\documentclass[12pt]{article}

\usepackage{amsfonts,amssymb,amsmath}
\usepackage{graphicx}
\usepackage{url}

\begin{document}

\title{\vspace{-2cm} {\normalsize Updated version to appear in the forthcoming Wiley Finance book \\ \bf{Counterparty Credit Risk,
Collateral and Funding}}
\\ ---  \\
{\bf\Large Counterparty Risk FAQ: \\{\large Credit VaR, PFE, CVA, DVA, Closeout, Netting, Collateral, Re-hypothecation, WWR, Basel, Funding, CCDS and Margin Lending}}
}
\author{
Damiano Brigo \\ {\normalsize Gilbart Professor of Financial Mathematics}\\ {\normalsize Head of the Financial Mathematics Research Group} \\ {\normalsize King's College, London. On move to Imperial College} \\ {\normalsize Paper available at  
damianobrigo.it \ \ defaultrisk.com \ \  ssrn.com \ \ arxiv.org}
}

\date{\small First Version: Oct 31, 2011. This Version: \today}

\maketitle

\vspace{-1cm}

\begin{abstract}
We present a dialogue on Counterparty Credit Risk touching on Credit Value at Risk (Credit VaR), Potential Future Exposure (PFE), Expected Exposure (EE), Expected Positive Exposure (EPE), Credit Valuation Adjustment (CVA), Debit Valuation Adjustment (DVA), DVA Hedging, Closeout conventions, Netting clauses, Collateral modeling, Gap Risk, Re-hypothecation, Wrong Way Risk, Basel III, inclusion of Funding costs, First to Default risk, Contingent Credit Default Swaps (CCDS) and CVA restructuring possibilities through margin lending. The dialogue is in the form of a Q\&A between a CVA expert and a newly hired colleague.
\end{abstract}
\noindent 
{\bf AMS classification codes: 60G51, 60G70, 60H35, 62G32, 65C05, 65C20, 91B70}


\newpage

\begin{center}
\includegraphics{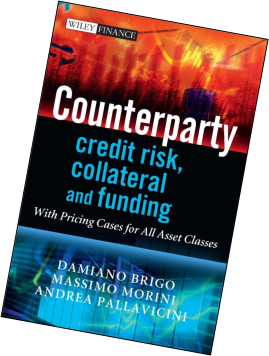}
\end{center}

\bigskip

\noindent {\Large An updated version of this dialogue will appear as the first chapter of the Wiley Finance (Nov 2012) Book

\bigskip

\noindent {\bf Counterparty Credit Risk,
Collateral}

{\bf  and Funding} with Pricing Cases for all Asset Classes

\bigskip

\noindent {\large by Damiano Brigo\footnote{This summary is based on my last 8 years of research, supported by several co-authors, including Claudio Albanese, Imane Bakkar, Tom Bielecki, Cristin Buescu, Agostino Capponi, Kyriakos Chourdakis, Massimo Morini, Frank Oertel, Andrea Pallavicini, Vasileios Papatheodorou, Frederic Patras, Daniele Perini and Roberto Torresetti. I am grateful to all participants in the dinner at the Royal Society in London on Oct 27, 2011, and to all participants of the Quant Congress Europe 2011, Global Derivatives 2012, Risk Minds US 2012, and CVA Collateral Damages 2012 for helpful discussions on CVA after my training courses and plenary talks. I am also grateful to Andrea Prampolini, Vasileios Papatheodorou and Cyril Durand for reading this document and providing helpful feedback. All remaining errors are my own. This paper expresses my opinion and is in no way representative of the institutions I work for or I am associated with.}, Massimo Morini and Andrea Pallavicini } 
}

\bigskip



{ \tableofcontents}

\pagestyle{myheadings}
\markboth{}{{\footnotesize D. Brigo: Counterparty Credit Risk FAQ / Dialogue}}

\newpage

\section{A Dialogue on CVA} 

Although research on counterparty risk pricing started way back in the nineties, with us joining the effort back in 2002, the different aspects of counterparty credit risk exploded after the beginning of the financial crisis in 2007. In less than four years we have seen the emergence of a number of features that the market operators are struggling to account for with consistency. Further, the several possible definitions and methodologies for counterparty risk may create further confusion. This dialogue is meant to provide a colloquial guide to the different aspects of counterparty risk. It is in the form of a Q\&A between a CVA expert and a newly hired colleague, and provides detailed references for investigating the different areas sketched here more in detail.

\section{Risk Measurement: Credit VaR}

\begin{itemize}
\item[Q:] [Junior colleague, he is looking a little worried] I am new in this area of counterparty risk, and I am struggling to understand the different measures and metrics. Could you start by explaining what is counterparty risk in general?

\item[A:] [Senior colleague, she is looking at the junior colleague reassuringly] The risk taken on by an entity entering an Over The Counter (OTC) contract
with one (or more) counterparty having a relevant default probability. As such,
the counterparty might not respect its payment obligations.

\item[Q:] What kind of counterparty risk practices are present in the market?

\item[A:] Several, but most can be divided into two broad areas. Counterparty risk measurement for capital requirements, following Basel II, or counterparty risk from a pricing point of view, when updating the price of instruments to account for possible default of the counterparty. However, the distinction is now fading with the advent of Basel III.

\item[Q:] [Shifts nervously] Let us disentangle this a little, I am getting confused. 

\item[A:] Fine. Where do we start from?

\item[Q:] Let us start from Counterparty Risk Measurement for Capital Requirements. What is that?

\item[A:] It is a risk that one bank faces in order to be able to lend money or invest towards a counterparty with relevant default risk. The bank needs to cover for that risk by setting Capital aside, and this can be done after the risk has been measured.

\item[Q:] You are saying that we aim at measuring that risk?

\item[A:] Indeed, and this measurement will help the bank decide how much capital the bank should set aside (capital requirement) in order to be able to face losses coming from possible defaults of counterparties the bank is dealing with.

\item[Q:] Could you make an example of such a measure?

 \item[A:] A popular such measure is Value at Risk (VaR). It is basically a percentile on the loss distribution associated with the position held by the bank, over a given time horizon. More precisely, it is a percentile (say the 99.9 percentile) of the initial value of the position minus the final value at the risk horizon, across scenarios.

\item[Q:] Which horizon is usually taken?

 \item[A:] When applied to default risk the horizon is usually one year and this is called ``Credit VaR". If this is taken at the 99.9-th percentile, then you have the loss that is exceeded only in 1 case out of 1000. The Credit VaR is actually either the difference of the percentile from the mean, or the percentile itself. There is more than one possible definition.

\item[Q:] Is this a good definition of Credit risk?

\item[A:] [Frowning] Well what does ``good" really mean? It is not a universally good measure. It has often been criticized, especially in the context of pure market risk without default, for lack of sub-additivity. In other terms, it does not always acknowledge the benefits of diversification, in that in some paradoxical situations the risk of a total portfolio can be larger than the sum of the risks in the single positions. A better measure from that point of view would be expected shortfall, known also as tail VaR, conditional VaR, etc.

\item[Q:] And what is that?

\item[A:] This is loosely defined as the expected value of the losses beyond the VaR point. But this needs not concern us too much here.

\item[Q:] Fine. How is Credit VaR typically calculated?

 \item[A:] Credit VaR is calculated through a simulation of the basic financial variables underlying the portfolio under the historical probability measure, commonly referred as $P$, up to the risk horizon. The simulation also includes the default of the counterparties. At the risk horizon, the portfolio is priced in every simulated scenario of the basic financial variables, including defaults, obtaining a number of scenarios for the portfolio value at the risk horizon.
 
\item[Q:] So if the risk horizon is one year, we obtain a number of scenarios for what will be the value of the portfolio in one year, based on the eveolution of the underlying market variables and on the possible default of the counterparties. 
 
 \item[A:] Precisely. A distribution of the losses of the portfolio is built based on these scenarios of portfolio values. When we say "priced" we mean to say that the discounted future cash flows of the portfolio after the risk horizon are averaged conditional on each scenario at the risk horizon but under another probability measure, the Pricing measure, or Risk Neutral measure, or Equivalent Martingale Measure if you want to go technical, commonly referred as $Q$.

\item[Q:] Not so clear... [Looks confused]

 \item[A:] [Sighing] All right, suppose your portfolio has a call option on equity, traded with a Corporate client, with a final maturity of two years. Suppose for simplicity there is no interest rate risk, so discounting is deterministic. To get the Credit-Var, roughly, you simulate the underlying equity under the $P$ measure up to one year, and obtain a number of scenarios for the underlying equity in one year. Also, you need to simulate the default scenarios up to one year, to know in each scenario whether the counterparties have defaulted or not. This default simulation up to one year is under the measure $P$ as well. And you may want to include the ``correlation" between default of the counterparty and underlying equity, that would allow you to model wrong way risk (WWR). But let us leave WWR aside for a moment. 
 
\item[Q:] Ok. We simulate under $P$ because we want the risk statistics of the portfolio in the real world, under the physical probability measure, and not under the so called pricing measure $Q$.   

\item[A:] That's right. And then in each scenario at one year, if the counterparty has defaulted there will be a recovery value and all else will be lost. Otherwise, we price the call option over the remaining year using for example a Black Scholes formula. But this price is like taking the expected value of the call option payoff in two years, conditional on each scenario for the underlying equity in one year. Because this is pricing, this expected value will be taken under the pricing measure $Q$, not $P$. This gives the Black Scholes formula if the underlying equity follows a geometric brownian motion under $Q$. 

\item[Q:] So default needs to be simulated only under $P$? Where do you find such probabilities?

\item[A:] [Frowning] This is a very difficult question. Often one uses probabilities obtained through aggregation, like the probability associated to the rating of the counterparty for example. But this is not very precise. Default of a single firm occurs only once, so determining the $P$ probability through direct historical observation is not possible...

\item[Q:] .....[shifts nervously on the chair].... 

\item[A:] [Concentrating] Notice also that, in a more refined valuation, you may want to take into account the default probability of the counterparty also between 1 and 2 years in valuing the call option. But this would be now the default probability under $Q$, not under $P$, because this is pricing. But let us leave this aside for the time being, because this leads directly to Credit Valuation Adjustments (CVA) which we will address later. It would be like saying that in one year you compute the option price value by taking into account its CVA. 

\item[Q:] [Frowning] I think I need to understand better this $P$ and $Q$ thing. For example, how are the default probabilities under $P$ and $Q$ different?

\item[A:] The ones under $Q$, typically inferred from market prices of CDS or corporate bonds, are typically larger than those under the measure $P$. This has been observed a number of times. A comparison of the $P$ and $Q$ loss distributions involved in Collateralized Debt Obligations (CDOs) is carried out for example in \cite{TorreJRMFI}.

\item[Q:] Some more acronyms... In the meantime, where can I read more about VaR and Expected Shortfall?

\item[A:] On a basic technical level you have books like \cite{jorion}, whereas to go at a higher technical level you have books like \cite{mcneil}. For the original Credit VaR framework it can be a good idea to have a look at the original ``Credit Metrics Technical Document" \cite{creditmetrics}, which is available for example at defaultrisk.com

\end{itemize}

\section{Exposure, CE, PFE, EPE, EE, EAD}

\begin{itemize}

\item[Q:] Ok, I have more or less understood Credit VaR and ES. But I also keep hearing the word ``Exposure" in a lot of meetings. What is that, precisely?

\item[A:] Let me borrow by \cite{CannDuff}. [Calls up a paper on the screen of her tablet]. These are not exactly the definitions and calculations used in Basel, we would need to go much more in detail for that,  but are enough to give you a good idea of what's going on.

\item[Q:] Hopefully... [looks at her senior colleague skeptically]

\item[A:] [Rolls her eyes] Counterparty exposure at any given future time is the larger between zero and the market value of the portfolio of derivative positions with a counterparty that would be lost if the counterparty were to default with zero recovery at that time. 

\item[Q:] This is clear. 

\item[A:] Current exposure (CE) is, obviously enough, the current value of the exposure to a counterparty. 
This is simply the current value of the portfolio if positive, and zero otherwise. This is typically the expected value under the pricing measure $Q$ of future cashflows, discounted back at the present time and added up, as seen from the present time, if positive, and zero otherwise. 

\item[Q:] Ok, I see.

\item[A:] Potential future exposure (PFE) for a given date is the maximum of exposure at that date with a high degree of statistical confidence. For example, the $95\%$ PFE is the level of potential exposure that is exceeded with only $5\%$ $P$-probability. The curve of PFE in time is the potential exposure profile, up to the final maturity of
the portfolio of trades with the counterparty. 

\item[Q:] Why 95? And what about $P$ and $Q$.

\item[A:] Just because [Amused]. On $P$ and $Q$, let's talk about that later.

\item[Q:] .... 

\item[A:] PFE is usually computed via simulation: for each future date, the price of the portfolio of trades with a counterparty is simulated. A $P$-percentile of the distribution of exposures is chosen to represent the
PFE at the future date. The peak of PFE over the life of the portfolio is called maximum potential
future exposure (MPFE). PFE and MPFE are usually compared with credit limits in
the process of permissioning trades.

\item[Q:] But wait... isn't this what you said about Credit VaR? Because earlier you said..

\item[A:] [Raising her hand] No, be careful... here there is no default simulation involved, only the portfolio is simulated but not the default of the counterparty. With exposure we answer the question: IF default happens, what is going to be the loss? 

\item[Q:] So in a way we assume that default happens for sure and we check what would be the loss in that case. I see. No default simulation or probabilities here.

\item[A:] Good. As we have seen above, with Credit VaR instead we answer the question: what is the final loss that is not exceeded with a  given $P$ probability, over a given time horizon? This second question obviously involves the inclusion of the default event of the counterparty in generating the loss. 

\item[Q:] Ok I understand. And that's it about Exposure, isnt't it? [Smiling hopefully]

\item[A:] By no means! [Amused]

\item[Q:] \begin{verbatim}@#$�!!@#&&��@@!!!!\end{verbatim} 
How many more bloody acronyms do I have to learn???

\item[A:] You should not use that language in a professional environment! 

\item[Q:] You are right, such expressions are unheard of in professional environments, I am behaving very poorly.

\item[A:] Do I detect a note of sarcasm? No matter. Here you go. Expected exposure (EE) is the average exposure under the $P$-measure on a future date. The curve of EE in time, as the future date varies, provides the expected exposure profile. Expected positive exposure (EPE) is the average EE in time up to a given future date (for example, for dates during a given year).

\item[Q:] Gosh... 

\item[A:] And did I mention Exposure at Default (EAD)?
This is simply defined as the exposure valued at the (random future) default time of the counterparty.  
 
\item[Q:] That's quite enough! [pulling his hair]

\end{itemize}

\section{Exposure and Credit VaR}

\begin{itemize}

\item[A:] [Looking at the junior colleague in a motherly fashion] Ok let's stop here. Basel II provided some rules and approximations explaining how such exposures could be approximated and calculated. Notice that the default probabilities are not part of this picture. There is no default simulation here, contrary to Credit VaR. 

\item[Q:] That's right, you never mentioned default modeling here. 

\item[A:] Essentially exposure measures how much you are likely to lose if the counterparty defaults. With Credit VaR we also add the default probability to the picture and get a final value for the possible loss inclusive of default probability information.  

\item[Q:] And why is exposure important?

\item[A:] Banks use to measure counterparty risk internally using mainly 2 measures: PFE, which is mainly used internally to monitor when the credit limits with the counterparties are breached, and EE, which is used, when combined with other quantities, for the calculation of EAD and the capital requirements due to counterparty risk. This last calculation may combine Exposures with default probabilities and recovery estimates, and it produces an approximation to Credit VaR, which is used as a capital requirement. 

\item[Q:] So we go back again to a percentile of the loss under a given risk horizon. What is the percentile and what is the risk horizon?

\item[A:] The risk horizon for this approximation of Credit VaR is typically one year and the confidence level is $99.9\%$ 

\item[Q:] That would seem to be quite safe

\item[A:] That seems safe, but the approximations and the assumptions introduced by Basel II to compute the approximated Credit VaR are not realistic and have been heavily critized. See the OECD paper \cite{oecd} for an overview of the problems, some of them also affecting Basel III. 

\end{itemize}

\section{Interlude: $P$ and $Q$}

\begin{itemize}

\item[Q:] More on $P$ and $Q$? You keep mentioning these two probabilities measures as if they were obvious, but I don't think they are... [looking worringly at his senior colleague] 

 \item[A:] [Frowning again] Statistical properties of random objects such as future losses depend on the probability measure we are using. Under two different probabilities a random variable will have usually two different expected values, variances, medians, modes, etc.

 \item[Q:] [Frowning in turn] So you are saying that a future random loss can have a different distribution under two different measures, such as $P$ and $Q$? But what is $P$ and what is $Q$, and why  do they differ?

 \item[A:] $P$, the historical or physical probability measure, also called real world probability measure, is the probability measure under which we do historical estimation of financial variables, econometrics, historical volatility estimation, maximum likelihood estimation, etc. When we need to simulate the financial variables up to the risk horizon we are using statistical techniques under $P$. When we try to make a prediction of future market variables, again, we do it under $P$.

 \item[Q:] I guess this is because prediction and risk measurement need to be done with the statistics of the observed world. But why introducing another probability measure $Q$? Why is it needed? [looking puzzled]

 \item[A:] If instead of simulating financial variables for prediction or risk measurement we are trying to price an option or a financial product, when we price products in a no-arbitrage framework, the no-arbitrage theory tells us that we need to take expected values of discounted future cash flows under a different probability measure, namely $Q$.

\item[Q:] And how is this $Q$ related to $P$? [still puzzled]

\item[A:]The two measures are related by a mathematical relationship that depends on risk aversion, or market price of risk. In the simplest models the real expected rate of return is given by the risk free rate plus the market price of risk times the volatility. Indeed the "expected" return of an asset depends on the probability measure that is used. For example, under $P$ the average rate of return of an asset is hard to estimate, whereas under $Q$ one knows that the rate of return will be the risk free rate, since dependence on the real rate of return can be hedged away through replication techniques. [starts looking tired]

\item[Q:] And why should this be interesting? [ironic]

\item[A:] Well, maybe it's not string theory or non-commutative topology (what did you say you studied for your PhD?), but the fact that arbitrage free theory removes uncertainty about the expected rate of return by substituting it with the risk free rate has been a big incentive in developing derivatives.

\item[Q:] Why is working under $P$ so difficult? [puzzled]

\item[A:] Determining the real world or $P$ expected return of an asset is difficult, and rightly so, or else we would all be rich by knowing good estimates of expected returns of all stocks in the future. [looks at the window dreamingly]

\item[A:] This is a lot to take in...

 \item[Q:]Let us say that you use $P$ until the risk horizon and then $Q$ to price the portfolio at the risk horizon.

\item[A:] I think I am starting to get a grip on this. So let me ask:
What is ``Basel''?

\item[Q:] A city in Switzerland?

\item[A:] Ah ah ah very funny...

\end{itemize}

\section{Basel}

\begin{itemize}

\item[A:]
%
Ok seriously... [pulls her tablet and visualizes a PDF document, handing the tablet to her junior colleague] ''Basel II'' is a set of recommendations on banking regulations issued by the Basel Committee on Banking Supervision. The ``II'' is due to the fact that this is a second set of rules, first issued in 2004 and updated later on, following a first set (Basel I) issued in 1998. Basel II has been introduced to  create a standard that regulators can use to establish how much capital a bank needs to set aside to cover for financial and operational risks connected to its lending and investing activities. Often banks tend to be willing to employ as much capital as possible, and so the more the reserves can be reduced while still covering the risks, the better for the banks. In other terms, often banks aim at reducing the capital requirements (i.e. the amounts to be set aside) to the minimum. Among Basel II purposes, the two most interesting for us are
\begin{itemize}
\item  Have capital requirements reflect more the risks and being more risk sensitive;
\item  Split operational risk and credit risk, quantifying both;
\end{itemize}
The capital requirements concern overall the three areas of credit - or \underline{counterparty} - risk, market risk, and operational risks. Here we deal with the first two mostly, and in particular with the first.
From this capital adequacy point of view, the counterparty risk component can be measured in three different frameworks of increasing complexity, the "standardized approach", the foundation Internal Rating-Based Approach (IRBA) and the advanced IRBA. The standardized approach employs conservative measures of capital requirements based on very simple calculations and quantities, so that if a bank follows that approach it is likely to find higher capital requirements than with the IRBA's. This is an incentive for banks to develop internal models for counterparty risk and credit rating, although the credit crisis started in 2007 is generating a lot of doubt and debate on the effectiveness of Basel II and of banking regulation more generally. Basel regulation is currently under revision in view of a new set of rules commonly referred to as Basel III. We will get to Basel III later.

\item[Q:] Is the Basel accord considered to be effective? Has there been any criticism?

\item[A:] You really are a rookie, aren't you? Of course there has been a lot of criticism. Have a look again at the OECD paper \cite{oecd}, for example.

\item[Q:] I'll do that. So, we mentioned above two broad areas: (i) Counterparty risk measurement for capital requirements, following Basel II, and the related Credit VaR risk measure, or (ii) counterparty risk from a pricing point of view. Basel II then concerns the capital one bank has to set aside in order to lend money or invest towards a counterparty with relevant default risk, to cover for that risk, and is related to Credit VaR. What about the other area, i.e. pricing?
\item[A:] Pricing concerns updating the value of a specific instrument or portfolio, traded with a counterparty, by altering the price to be charged to the counterparty. This modification in price is done to account for the default risk of the counterparty. Clearly, all things being equal, we would always prefer entering a trade with a default-free counterparty than with a default risky one. Therefore we charge the default risky one a supplementary amount besides the default-free cost of the contract. This is often called Credit Valuation Adjustment, or CVA. Since it is a price, it is computed entirely under the $Q$ probability measure, the pricing measure. In principle, the $P$ probability measure does not play a role here. We are computing a price, not measuring risk statistics.

\item[Q:] Has this concept been there for a long time or is it recent?

\item[A:] It has been there for a while, see for example \cite{DuffieHuang}, \cite{BielRut}, \cite{BrigoMas}. However it became more and more important after the 2008 defaults.

\end{itemize}

\section{CVA and Model Dependence}

\begin{itemize}

\item[Q:] But this CVA term, how does it look like?

 \item[A:] It looks like an option on the residual value of the portfolio, with a random maturity given by the default time of the counterparty.

\item[Q:] Why an option? How does it originate?

 \item[A:] If the counterparty defaults and the present value of the portfolio at default is positive to the surviving party, then the surviving party only gets a recovery fraction of the portfolio value from the defaulted entity. If however the present value is negative to the surviving party, the surviving party has to pay it in full to the liquidators of the defaulted entity. This creates and asymmetry that, once one has done all calculations, says that the value of the deal under counterparty risk is the value without counterparty risk minus a positive adjustment, called CVA. This adjustment is the price of an option in the above sense. See again \cite{BrigoMas} for details and a discussion.

\item[Q:] A price of an option with random maturity? Looks like a complicated object... [frowning]

\item[A:] [Smiling] It is, and it is good that you realize it. Indeed it is quite complicated. First of all, this is complicated because it introduces model dependence even in products that were model independent to start with. Take for example a portfolio of plain vanilla swaps. You don't need a dynamic term structure model to price those, but only the curves at the initial time.

\item[Q:] And what happens with CVA?

 \item[A:] Now you have to price an  option on the residual value of the portfolio at default of the counterparty. To price an option on a swap portfolio you need an interest rate option model. Therefore even if you portfolio valuation was model independent before including counterparty risk, now it is model dependent. This means that quick fixes to pricing libraries are quite difficult to obtain.

\item[Q:] I see... Model dependence... and model risk. So anyway volatilities and correlations would impact this calculation?

 \item[A:] Yes, and dynamics features more generally. Volatilities of the underlying portfolio variables and also of the counterparty credit spreads all impact valuation importantly. But also the statistical dependence (or "correlation") between default of the counterparty and underlying financial variables, leading to so called wrong way risk, can be very important.

\item[Q:] Wrong Way Risk? WWR?

\item[A:] Yes, I am sure you heard this before.

\item[Q:] Well I am not sure about WWR, but before we go there hold on a minute, I have another question.

\item[A:] [Sighing] Go ahead.

\end{itemize}

\section{Input and Data issues on CVA}

\begin{itemize}

\item[Q:] You mentioned volatilities a correlations, but are they easy to measure?

\item[A:] That is both a very good and important question. No, they are not easy to measure. We are pricing under the measure $Q$, so we would need volatilities and correlation extracted from traded prices of products that depend on such parameters.

\item[Q:] But where can I extract the correlation between a specific corporate counterparty default and the underlying of the trade, for example oil, or a specific FX rate? And where do I extract credit spread volatilities from?

\item[A:] [Looks at the young colleage with increased attention] You are not a rookie then if you ask such questions, you must have some experience.

\item[Q:] [Sighing] Not really... I heard such questions at a meeting of the new products committee yesterday, as I was sitting in a corner as the resident newbie, and started thinking about these issues.

\item[A:] [Sighing in turn] Well at least you learn fast. Let me tell you that the situation is actually worse. For some counterparties it is even difficult to find levels for their default probabilities, not to mention expected recoveries.

\item[Q:] Aren't $Q$ default probabilities deduced from Credit Default Swap or corporate bond counterparty data?

\item[A:] Yes they are... in principle. But many counterparties do not have a liquid CDS or even bond traded. What if your counterparty is the airport of Duckburg? Where are you going to imply default probabilities from, leave alone credit volatilities and credit-underlying "correlations"? And recoveries?

\item[Q:] Recoveries, indeed, aren't those just 0.4? [Grinning]

\item[A:] [Rolling eyes] Right. Just like that. However let me mention that when the $Q$ statistics are not available, one first attempt one can consider is using $P$-statistics instead. One can estimate credit spread volatility historically if no CDS or corporate bond option implied volatility is available. Also historical correlations between the counterparty credit spreads and the underlying portfolio of the trade can be much easier to access than implied ones. It is clearly an approximation but it is better than no idea at all. Even default probabilities, when not available under $Q$, may be considered under $P$ and then perhaps ajusted for an aggregate estimate of credit risk premia. Rating information can provide rough aggregate default probabilities for entities such as the airport of Duckburg if one has either an internal or external rating for small medium enterprises (SME).

\item[Q:] Aren't there a lot of problems with rating agencies?

\item[A:] Yes there are, and I am open to better ideas if you have anything to propose.

\item[Q:] Not easy... But leaving aside default probabilities, credit correlations, credit-underlying correlations, and recoveries... 

\item[A:] You are leaving aside quite some material...

\item[Q:] ... what about the underlying contract $Q$ dynamics, is that clear for all asset classes? 

\item[A:] For a number of asset classes, traditional derivatives markets provide you with underlying market levels, volatilities and market-market "correlations". But not always.  

\item[Q:] Can you provide an example where this does not work?

\end{itemize}

\section{Emerging asset classes: Longevity Risk}

\begin{itemize}

\item[A:] Let me think... yes that could be a good example, longevity risk.

\item[Q:] I was never sure how to pronounce that in English.

\item[A:] It is longevity, [lon-jev-i-tee], as I pronounced it, "ge" like in "George" rather than "get". 

\item[Q:] Longevity... but what kind of risk is that? I wouldn't mind leaving a long time, provided the quality of life is good. 

\item[A:] It is not a risk for you, it is a risk for your pension provider. If you leave longer than expected then the pension fund needs extra funding to keep your pension going. 

\item[Q:] Right [touching wood on the table]. 

\item[A:] [Laughing] If you find the name disturbing, we may call it mortality risk. Anyway with longevity swaps the problem is also finding the underlying $Q$-dynamics, both in levels and volatilities, namely levels and volatilities of mortality rates...

\item[Q:] Wait a minute. Longevity {\emph{swaps}}? What is a longevity swap? Sounds like a pact with the devil for longer life in exchange of your soul or...

\item[A:] [Raising her hand] Can't you be professional for a minute? A longevity swap is a contract where one party (typically a pension fund) pays a pre-assigned interest rate in exchange for a floating rate linked to the realized mortality rate in a given country, or area, over a past window of time.

\item[Q:] Sorry for the interruptions, ok this makes sense. So I guess the problem is the calibration of the mortality rate dynamics in pricing the future cash flows of the swap?

\item[A:] Indeed, the problem is that for this product there is basically almost no information from which one can deduce the $Q$ dynamics... I wonder actually if it even makes sense to talk about $Q$-dynamics. If swaps were very liquid we could imply a term structure of mortality rates from the prices, and also possibly implied volatilities if options on this swaps became liquid. 

\item[Q:] And I imagine that, being linked to pensions, these contracts have large maturities, so that counterparty risk is relevant? 

\item[A:] Precisely. Now this is an emerging area for counterparty risk, with almost no literature, except the excellent initial paper \cite{biffisblake}. In terms of $Q$-dynamics, a first approach could be to use the $P$-dynamics and assume there is no market price of risk, at least until the market develops a little further. 

\item[Q:] Here I think it may be really hard to find data for the statistical dependence between the underlying mortality rates and the default of the counterparty, which brings us back to the subject of Wrong Way Risk, on which I have many general questions.

\end{itemize}

\section{CVA and Wrong Way Risk}

\begin{itemize}

\item[A:] [Shifting on the chair] Oh I'm sure you do! Let me try and anticipate a few of them. WWR is the additional risk you have when the underlying portfolio and the default of the counterparty are ``correlated" in  the worst possible way for you. 

\item[Q:] For example?

\item[A:] Suppose you are trading an oil swap with an airline and you are receiving floating (variable) oil and paying fixed. We may envisage a positive correlation between the default of the airline and the price of oil, since higher prices of oil will put the airline under more stress to finance her operations. When the correlation is extremely high, so that at a marked increase of oil there is a corresponding marked increased in the airline default probability, we have the worst possible loss at default of the airline. Indeed, with high oil price increase the oil swap now has a much larger value for us, and there is a higher probability of default of the airline due to the correlation. If the airline defaults now it will do so in a state where the mark to market is quite high in our favor, so that we face a large loss.   This is an example of wrong way risk.

\item[Q:] Has Wrong Way Risk been studied?

\item[A:] Yes, see for example the following references for such issues in different asset classes: \cite{BrigoMas}, \cite{BrigoMoriniTarenghi}, \cite{BrigoTarenghi2004}, \cite{BrigoTarenghi2005} for equity, \cite{BrigoPalla07}, \cite{BrigoPalla08} for interest rates, \cite{BrigoChourBakkar} for commodities (Oil),  \cite{Brigo08} for Credit (CDS).

\item[Q:] So there has been literature available on wrong way risk. Going back to the option structure of CVA, since options are priced under $Q$, I would guess that CVA calculations occur mostly under $Q$. But can one really work only under $Q$?

\item[A:] Before the crisis started in 2007, in front office environment it has been relatively common to work under $Q$, forgetting about $P$. One would postulate models for market processes and then calibrate them to prices, that are expectations under $Q$. Then simulations to compute prices of other products as expected values would still be done under $Q$. Similarly, to compute hedge ratios $Q$ used to be enough. $P$ used to be ignored except for risk measurement and possibly stress testing and model validation.

\item[Q:] And is this a good thing? [perlexed]

\item[A:] [Frowning] It is good because it allows you to avoid modeling the same processes under two probability measures, which could be rather tricky, since the real world $P$ statistics are often hard to obtain, as we explained above. But on the other hand one should really do a combined estimation of a pricing model based on the observed history of prices. The prices are $Q$ expectations but they move in time following the evolution of basic market variables under the $P$ measure. Kalman and more generally non-linear filtering techniques can be used to obtain a joint estimation of the underlying market processes, which would incorporate market history ($P$) AND risk neutral expectations ($Q$) at the same time. This implicitly estimates also market aversion, connecting $P$ and $Q$.

\item[Q:] So all the attention to Counterparty risk now is about $P$ (Credit VaR) or $Q$ (CVA)?

\item[A:] [Looking at the ceiling] At the moment most attention is on CVA, but now with Basel III the distinction is blurring.

\end{itemize}

\section{Basel III: VaR of CVA and Wrong Way Risk}

\begin{itemize}

\item[Q:] What do you mean? Give me a break! It is already complicated enough!

\item[A:] Relax. Let us say that Credit VaR measures the risk of losses you face due to the possible default of some counterparties you are having business with. CVA measures the pricing component of this risk, i.e. the adjustment to the price of a product due to this risk.

\item[Q:] This is clear.

\item[A:] But now suppose that you revalue and mark to market CVA in time. Suppose that CVA moves in time and moves against you, so that you have to book negative losses NOT because the counterparty actually defaults, but because the Pricing of this risk has changed for the worse for you. So in this sense you are being affected by CVA volatility.

\item[Q:] Ah...

\item[A:] To quote Basel III: [Visualizes a document on her tablet]

``Under Basel II, the risk of counterparty default and credit migration risk were addressed but mark-to-market losses due to credit valuation adjustments (CVA) were not. During the financial crisis, however, roughly two-thirds of losses attributed to counterparty credit risk were due to CVA losses and only about one-third were due to actual defaults."

\item[Q:] So in a way the variability of the price of this risk over time has made more damage than the risk itself?

\item[A:] I guess you could put it that way, yes. This is why Basel is considering setting up quite severe capital charges against CVA.

\item[Q:] And why did you say that this blurs the picture?

\item[A:] Because, now, you may decide that you need a VaR estimate for your CVA, especially after the above Basel III statement.

\item[Q:] How would this be computed?

\item[A:] You could simulate basic market variables under $P$, up to the risk horizon. Then, in each scenario, you price the residual CVA until the final maturity using a $Q$ expectation. You put all the prices at the horizon time together in a histogram and obtain a profit and loss distribution for CVA at the risk horizon. On this $P$ distribution you select a quantile at the chosen confidence level and now you will have computed VaR of CVA. But this does not measure the default risk directly, it measures the risk to have a mark to market loss due either to default or to adverse CVA change in value over time.

 \item[Q:]... while Credit VaR only measures the default risk, i.e. the risk of a loss due to a direct default of the counterparty.
Let's go back to counterparty risk as a whole now. Where is our focus in all of this?

 \item[A:] Here we are dealing mostly with CVA valuation. So we give more relevance to $Q$ than $P$, but we'll have a number of comments on $P$ as well.

\item[Q:] So what is Basel III saying about CVA, specifically?

 \item[A:] Well the framework has changed several times: Bond Equivalent formula, multipliers... One of the main issues has to do with Wrong Way Risk (WWR).

\item[Q:] What do you mean?

 \item[A:] In some part of the Basel regulation it had been argued that you could calculate CVA as if there were no wrong way risk, and then use a standard multiplier to account for wrong way risk.

 \item[Q:] So I should assume independence between default of the counterparty and underlying portfolio, compute CVA, and then multiply for a given number to account for correlation risk?

\item[A:] Something like that. However, this does not work. Depending on the specific dynamics of the underlying financial variables, on volatilities and correlations, and on the chosen models, the multipliers vary a lot. See again \cite{BrigoMoriniTarenghi}, \cite{BrigoPalla07}, \cite{BrigoChourBakkar} and \cite{Brigo08} for examples from several asset classes. The multipliers are very volatile and fixing them is not a good idea. Even if one were to use this idea only for setting capital requirements of well diversified portfolios, this could lead to bitter surprises in situations of systemic risk. And there is a further problem...

\item[Q:] What else???!!? [looking desperate]

\end{itemize}

\section{Discrepancies in CVA valuation: Model risk and Payoff Risk}

\begin{itemize}

\item[A:] Relax, relax... Look we can take a break, you look too distressed.
 
\item[Q:] Ok let's have a coffee.

\item[A:] I would advise a camomille...

\bigskip

[Twenty minutes later]

\bigskip
 
\item[A:] Let's go on. Basel III recognizes CVA risk but does not recognize DVA risk, the quantity one needs to introduce to make counterparty risk work from an accounting perspective. This creates a misalignment between CVA calculations for capital adequacy purposes and CVA calculations for accounting and mark to market. This is part of a more general problem.

\item[Q:] Really? Sounds unbelievable that two bits of regulation can be at odds like that!

\item[A:] I can show you. Look at this:

\bigskip
”This CVA loss is calculated without taking into account any offsetting
debit valuation adjustments which have been deducted from capital
under paragraph 75.”
(Basel III, page 37, July 2011 release)

\bigskip
”Because nonperformance risk (the risk that the obligation will not be
fulfilled) includes the reporting entity’s credit risk, the reporting entity
should consider the effect of its credit risk (credit standing) on the fair
value of the liability in all periods in which the liability is measured at
fair value under other accounting pronouncements" (FAS 157) 

\item[Q:] Surprisingly clear and at odds. 

\item[A:] Well since you seem to enjoy it, here is what the former president of the Basel Committee said: 

\bigskip

"The potential for perverse incentives resulting from profit being linked
to decreasing creditworthiness means capital requirements cannot
recognise it, says Stefan Walter, {\em secretary-general of the Basel
Committee}: The main reason for not recognising DVA as an offset is
that it would be inconsistent with the overarching supervisory prudence
principle under which we do not give credit for increases in regulatory
capital arising from a deterioration in the firms own credit quality."
(Stefan Walter)

\item[Q:] I am quite confused. Should I compute DVA or not?

\item[A:] It depends on the purpose you are computing it for. However, the situation is not that clear, there are a number of issues more generally with counterparty risk pricing. 

\item[Q:] You mean objectivity on CVA valuation?

\item[A:] I mean that there is a lot of model risk and of ``payoff risk" if we want to call it that.

\item[Q:] I understand model risk, since this is highly model dependent, but what do you mean by payoff risk?

\item[A:] There are a lot of choices to be done when computing CVA, both on the models to be used, and on the type of CVA to be computed. We will see below that there are choices to be made on whether it is unilateral or bilateral, on the closeout formulation, on how you account for collateral and re-hypothecation, on whether you include first to default, and on how you account for funding costs, and so on. Due to the variety of possible different definitions of CVA and of modeling choices, there appears to be material discrepancies in CVA valuation across financial institutions, as pointed out in the recent article \cite{watt}.

\end{itemize}

\section{Bilateral Counterparty Risk: CVA and DVA}

\begin{itemize}

\item[Q:] Wait you're going too fast. You mentioned DVA above and I don't even know what it is. What is DVA?

\item[A:] Debit Valuation Adjustment. It has to do with both parties in a deal agreeing on the counterparty risk charge.

\item[Q:] Let me get this straight. Let's say that we are doing pricing, at a point in time, of the risk that the counterparty defaults before the final maturity of the deal, on a specific portfolio. This is the CVA. It is a positive quantity, an adjustment to be subtracted from the default-risk free price in order to account for the counterparty default risk in the valuation. Clearly, having the choice and all things being equal, one would prefer to trade a deal with a default risk free counterparty rather than with a risky one. So I understand the risk free price needs to be decreased through a negative adjustment, i.e. the subtraction of a positive term  called CVA. Now you are implicitly raising the question: Since it is an adjustment and it is always negative, what happens from the point of view of the other party?

\item[A:] Indeed, that's what I am saying. In this setup there is no possibility for both parties to agree, unless they both recognize that one of the calculating parties is default free. Suppose we have two parties in the deal, a bank and a corporate counterparty. If they both agree that the bank can be treated as default-free, then the bank will mark a negative adjustment on the risk free price of the deal with the corporate client, and the corporate client will mark a corresponding positive adjustment (the opposite of the negative one) to the risk free price. This way both parties will agree on the price.

\item[Q:] The adjustment for the corporate client is positive because the client needs to compensate the bank for the client default risk?

\item[A:] Indeed, this is the case. The adjustment seen from the point of view of the corporate client is positive, and is called Debit Valuation Adjustment, DVA. It is positive because the early default of the client itself would imply a discount on the client payment obligations, and this means a gain in a way. So the client marks a positive adjustment over the risk free price by adding the positive amount called DVA. In this case, where the bank is default free, the DVA is also called Unilateral DVA, UDVA, since only the default risk of the client is included. Similarly, the adjustment marked by the bank by subtraction is called Unilateral CVA, UCVA. In this case UCVA(bank) = UDVA(corporate), ie the adjustment to the risk free price is the same, but it is added by the corporate client and subtracted by the bank.

\item[Q:] But then the UCVA(corporate) must be zero, because the bank is default free.

\item[Q:] Correct, and similarly UDVA(bank) = UCVA(corporate) = 0.

\item[Q:] But what happens when the two firms do not agree on one of them being default free? Say that in your example the corporate client does not accept the bank as default free (a reasonable objection after Lehman...)

\item[A:] Well in this case then the only possibility to agree on a price is for both parties to consistently include both defaults into the valuation. Hence every party needs to include its own default besides the default of the counterparty into the valuation. Now both parties will mark a positive CVA to be subtracted and a positive DVA to be added to the default risk free price of the deal. The CVA of one party will be the DVA of the other one and viceversa.

\item[Q:] So every party will compute the final price as [writes on a notebook]

DEFAULT RISK FREE PRICE + DVA - CVA ?

\item[A:] Indeed. In our example when the bank does the calculation,

Price To Bank = DEFAULT RISK FREE PRICE to Bank + DVA Bank - CVA Bank

whereas when the corporate does the calculation one has a similar formula. Now, since

DEFAULT RISK FREE PRICE to Bank = - DEFAULT RISK FREE PRICE to Corporate

DVA Bank = CVA Corporate

DVA Corporate = CVA Bank

we get that eventually

Price To Bank = - Price To Corporate

so that both parties agree on the price, or, we could say, there is money conservation.

We could call Bilateral Valuation Adjustment (BVA) to one party the difference DVA - CVA as seen  from that party,

BVA = DVA - CVA

Clearly BVA to Bank = - BVA to corporate.

\item[Q:] Clear enough... so what is meant usually by ``bilateral CVA"?

\item[A:] Good question. By looking at the formula

BVA = DVA - CVA

bilateral CVA could refer both to BVA, or just to the CVA component of BVA on the right hand side. Usually the industry uses the term to denote BVA, and we will do so similarly, except when explicitly countered.

\item[Q:] Ok, summarizing... if we ask when valuation of counterparty risk is symmetric, meaning that if the other party computes the counterparty risk adjustment towards us she finds the opposite number, so that both parties agree on the charge, the answer is... [hesitating]

\item[A:] The answer is that this happens when we include the possibility that also the entity
computing the counterparty risk adjustment (i.e. us in the above example) may default, besides
the counterparty itself.

\item[Q:] Is there any technical literature on Bilateral CVA and on DVA?

 \item[A:]Yes, the first calculations are probably again in \cite{DuffieHuang}, who however resort to specific modeling choices where credit risk is purely accounted for by spreads and it is hard to create a strong dependence between underlying and default, so that Wrong Way Risk is hard to model. Furthermore, that paper deals mostly with swaps. Again, swaps with bilateral default risk are dealt with in \cite{BielRut}, but the paper where bilateral risk is examined in detail and DVA derived is \cite{BrigoCapponi}, where bilateral risk is introduced in general and then analyzed for CDS. In the following works \cite{BrigoPallaPapa}, \cite{BrigoCapponiPalla}, and \cite{brigoetalcollateral} several other features of bilateral risk are carefully examined, also in relationship with wrong way risk, collateral and extreme contagion, and gap risk, showing up when default happens between margining dates and a relevant mark to market change for the worse has occurred. In this respect, \cite{BrigoCapponiPalla} shows a case of an underlying CDS with strong default contagion where even frequent margining in collateralization is quite uneffective. For a basic introduction to bilateral CVA see \cite{Gregory}.

\item[Q:] There's too much material to read already!

\item[A:] Well that's why I'm trying to give you a summary here.

\item[Q:] Ok thanks, at least now I have an idea of what Bilateral CVA and DVA are about.

\end{itemize}

\section{First to Default in CVA and DVA}

\begin{itemize}

\item[A:] Yes, but you have to be careful. BVA is not just the difference of DVA and CVA computed each as if in a universe where only one name can default. In computing DVA and CVA in the difference you need to account for both defaults of Bank and Corporate in both terms. This means that effectively there is a first to default check. If the bank is doing the calculations, in scenarios where the bank defaults first the DVA term will be activated and the CVA term vanishes, whereas in scenarios where the corporate defaults first then the bank DVA vanishes and the bank CVA payoff activates. So we need to check who defaults first.

\item[Q:] Indeed, I heard ``first to default risk" in connection with Bilateral CVA.
Now, in computing the CVA and DVA terms, we should know who defaults first, that's what you are saying. That makes sense: to close the position in the right way and at the right time, I need to know who defaults first and when.

\item[A:] Correct. However, some practitioners implemented a version of BVA that ignores first to default times. Suppose you are the bank. Then for you

BVABank = DVABank - CVABank

See for example \cite{Picoult2005}. What you do now is computing DVABank as in a world where only you may default, and then compute CVABank as in a world where only the corporate client may default. But you do not kill the other term as soon as there is a first default. So in a sense you are double counting, because you are not really closing the deal at the first default if you do as we just said. The correct BVA includes a first to default check.

\item[Q:] My head is spinning... let me try to summarize.

\item[A:] Go ahead

\item[Q:] You have to be careful with Bilateral CVA. BVA is not just the difference of DVA and CVA computed each as if in a universe where only one name can default. In computing DVA and CVA in the difference you need to account for both defaults of Bank and Corporate \emph{in both terms}. This means that effectively there is a first to default check. If the bank is doing the calculations, in scenarios where the bank defaults first the DVA term will be activated and the CVA term vanishes, whereas in scenarios where the corporate defaults first then the bank DVA vanishes and the bank CVA payoff activates. So we need to check who defaults first.

\item[A:] Excellent, even better than my original explanation. More than a summary it looks like an essay!

\item[Q:] Well, not that I am going to write a paper on this.

\item[A:] Someone did already, see \cite{BrigoBuescuMorini}. The error in neglecting the first to default risk can be quite sizeable even in simple examples.

\end{itemize}

\section{DVA mark to market and DVA hedging}

\begin{itemize}

\item[Q:] I don't know, from all you told me I am not at ease with this idea of DVA. It is a reduction on my debt due to the fact that I may default, and if I default I won't pay all my debt, so it is like a gain, but I only can realize this gain as a cash flow if I default!!

\item[A:] I agree it can be disconcerting. And consider this: if your credit quality worsens and you recompute your DVA, you mark a gain.

\item[Q:] Has this really happened?

\item[A:] Citigroup in its press release on the first quarter revenues of 2009 reported a {\em positive} mark to market due to its {\em worsened} credit quality: [pulls out her tablet] ``Revenues also included [...] a net 2.5\$ billion positive CVA on derivative positions, excluding monolines, mainly due to the widening of Citi's CDS spreads"

\item[Q:] Ah...

\item[A:]
More recently, From the Wall Street Journal

{\tt October 18, 2011, 3:59 PM ET. Goldman Sachs Hedges Its Way to Less Volatile Earnings.}
Goldman's DVA gains in the third quarter totaled \$450 million, about \$300 million of which was recorded under its fixed-income, currency and commodities trading segment and another \$150 million recorded under equities trading.
That amount is comparatively smaller than the \$1.9 billion in DVA gains that J.P. Morgan Chase and Citigroup each recorded for the third quarter. Bank of America reported \$1.7 billion of DVA gains in its investment bank.
Analysts estimated that Morgan Stanley will record \$1.5 billion of net DVA gains when it reports earnings on Wednesday
[...]

\item[Q:] Sounds strange, you gain from the deterioration of your credit quality... and you lose from improvement of your credit quality. So how could DVA be hedged? One should sell protection on oneself, an impossible feat, unless one buys back bonds that he had issued earlier. This may be hard to implement, though. [looking more and more puzzled]

\item[A:]
It seems that most times DVA is hedged by proxying. Instead of selling protection on oneself, one sells protection on a number of names that one thinks are highly correlated to oneself.
Again from the WSJ article above: [toting her tablet at the junior colleague]

``[...] Goldman Sachs CFO David Viniar said Tuesday that the company attempts to hedge [DVA] using a basket of different financials.
A Goldman spokesman confirmed that the company did this by selling CDS on a range of financial firms.
[...]
Goldman wouldn�t say what specific financials were in the basket, but Viniar confirmed [...] that the basket contained �a peer group.� Most would consider peers to Goldman to be other large banks with big investment-banking divisions, including Morgan Stanley, J.P. Morgan Chase, Bank of America, Citigroup and others. The performance of these companies� bonds would be highly correlated to Goldman's."

\item[Q:] It seems to be relatively common practice then. Mmmmmm... isn't it risky? Proxying can be misleading [shaking his head]

\item[A:]
[Shrugging] Admittedly... This can approximately hedge the spread risk of DVA, but not the jump to default risk. Merrill hedging DVA risk by selling protection on Lehman would not have been a good idea. In fact this attitude, in presence of jump to default risk, can worsen systemic risk.

\item[Q:] Indeed, I can see that. If I sell protection on a firm that is correlated to me to hedge my DVA, and then that firm not only has his credit quality worsen (which would hedge my DVA changes due to spread movements) but actually defaults, then I have to make the protection payments and, paradoxycally, that could push me into default! [Looking at the senior colleague excitedly] 

\item[A:]
Sounds crazy, isn't it? [Grinning]  

\end{itemize}

\section{Impact of Closeout in CVA and DVA}

\begin{itemize}

\item[Q:] Well, to be perfectly honest, it does, but maybe it's just you and I being not sophisticated enough. However, there seem to be other matters that are as pressing.
I am having a hard time figuring out what this other problem with closeout is, for example.

\item[A:] [Sighs] Closeout is what happens basically when one name defaults. So suppose in our example the corporate client defaults. Closeout proceedings are then started according to regulations and ISDA\footnote{International Swaps and Derivatives Association} documentation. The closeout procedure establishes the residual value of the contract to the bank, and how much of that is going to be paid to the bank party, provided it it positive. If it is negative then the bank will have to pay the whole amount to the corporate.

\item[Q:] Well this seems simply the definition of CVA payout.

\item[A:] Ah, but let me ask you a question. At the default time of the Corporate, you are the Bank. Do you value the remaining contract by taking into account your own credit quality (in other terms, your now unilateral DVA, ``replacement closeout") or just by using the risk free price (``risk free closeout")?
The replacement closeout argues that if you are going now to re-open the deal with a risk free party, the risk free party will charge you your unilateral CVA, which, seen from your point of view, is your unilateral DVA. Hence in computing the replacement value you should include your DVA to avoid discontinuity in valuation. If you always used DVA to value the deal prior to the Corporate's default, you should not stop doing so at default  if you aim at being consistent.

\item[Q:] But there seem to be two choices here, risk free or replacement closeout. What is the difference? Is it just consistency and continuity of valuation?

\item[A:] The counterparty risk adjustments change strongly
depending on which assumption is chosen in the computation of the closeout
amount, and the choice has important consequences on default contagion.

\item[Q:] I would naively think that risk free closeout is simpler and more ``objective".

\item[A:] Well in \cite{BrigoMorini2011Risk}, \cite{BrigoMorini2010} and \cite{BrigoMorini2010Flux} it is shown that a risk-free closeout has implications that are very different from what we are used to expect in case of a
default in standardized markets such as the bond or loan markets. Let us take a case of BVA where the valuation is always in the same direction, such as a loan or a bond. Suppose the bank owns the bond. If the owner of a bond defaults, or if the lender in a loan defaults, this means no
losses to the bond issuer (the Corporate in our example) or to the loan borrower. Instead, if the risk-free
default closeout applies, when there is default of the party which is a net
creditor in a derivative (thus in a position similar to a bond owner or loan
lender, the Bank), the value of the liability of the net debtor will suddenly jump up.
In fact, before the default, the liability of the net debtor had a
mark-to-market that took into account the risk of default of the debtor
itself. After the default of the creditor, if a risk-free closeout applies,
this mark-to-market transforms into a risk-free one, surely larger in
absolute value than the pre-default mark-to-market.

\item[Q:] This appears to be definitely wrong. [Shaking his head again]

\end{itemize}

\section{Closeout Contagion}

\begin{itemize}

\item[A:] You are taking it too personally. Calm down.
It's actually worse.  The increase will be larger the larger the credit spreads of the debtor.
This is a dramatic surprise for the debtor that will soon have to pay this
increased amount of money to the liquidators of the defaulted party. There
is a true contagion of a default event towards the debtors of a defaulted
entity, that does not exist in the bond or loan market. Net debtors at
default will not like a risk-free closeout. They will prefer a replacement
closeout, which does not imply a necessary increase of the liabilities since
it continues taking into account the credit-worthiness of the debtor also
after the default of the creditor.

\item[Q:] You are saying that the replacement
closeout inherits one property typical of fundamental markets: if one of the
two parties in the deal has no future obligations, like a bond or option
holder, his default probability does not influence the value of the deal at
inception.

\item[A:] Correct. One could, based on this, decide to use replacement closeout all the time, since it is consistent with this basic principle.
However, the replacement closeout has shortcomings opposite to
those of the risk-free closeout. While the replacement closeout is preferred
by debtors of a defaulted company, symmetrically a risk-free closeout will
be preferred by the creditors. The more money debtors pay, the higher the
recovery will be. The replacement closeout, while protecting debtors, can in
some situations worryingly penalize the creditors by abating the recovery.

\item[Q:] What are such cases?

\item[A:] Consider the case when the defaulted entity is a company with high systemic
impact, so that when it defaults the credit spreads of its counterparties
are expected to jump high. Lehman's default could be a good example of such
a situation. If the credit spreads of the counterparties increase at
default, under a replacement closeout the market value of their liabilities
will be strongly reduced, since it will take into account the reduced
credit-worthiness of the debtors themselves. All the claims of the
liquidators towards the debtors of the defaulted company will be deflated,
and the low level of the recovery may be again a dramatic surprise, but this
time for the creditors of the defaulted company.

\item[Q:] [Baffled] It seems unbelievable that no clear regulation was available for this issue.

\item[A:] [Sighing] Well this is because there is no ideal solution. You may summarize the choice according to this table, let me draw it for you [Draws Table \ref{tab:contagioncloseoutintro} on her tablet]. As you see, there is no optimal choice guaranteeing no contagion. Depending on the ``correlation" structure between default of the borrower and the lender party in the transaction, the optimal choice is different. ISDA cannot set a standard that is correlation dependent, so it is understandable that there are difficulties in standardizing closeout issues.

\begin{table}[h!]
\begin{tabular}{|c|cc|}
\hline
\mbox{Dependence}$\rightarrow $ & \mbox{independence} & %
\mbox{co-monotonicity} \\
\mbox{Closeout}$\downarrow $ &  &  \\ \hline
\mbox{Risk Free} & Negatively affects & No contagion \\
& Borrower &  \\
&  &  \\
\mbox{Replacement} & No contagion & Further Negatively \\
&  & affects Lender \\ \hline
\end{tabular}%
\caption{Impact of the default of the Lender (Bank) under risk free or replacement closeout and under independence or co-monotonicity between default of the Lender and of the Borrower (Corporate)}\label{tab:contagioncloseoutintro}
\end{table}

\item[Q:] It looks more and more complicated. So many choices...

\item[A:] It's not over yet in terms of issues. But it's not that bad, that will keep us working for a long time [sarcastically].

\item[Q:] So what are the next issues that are keeping CVA people busy?

\item[A:] Collateral modeling, possible re-hypothecation, netting, Capital requirements around CVA for Basel III and possibilities to reduce them through restructuring, collateral or margin lending. Finally, consistent inclusion of funding costs...

\item[Q:] [Rolling his eyes] That's quite enough. Let's start from Collateral.

\end{itemize}

\section{Collateral Modeling in CVA and DVA}

\begin{itemize}

 \item[A:]Collateral is an asset (say Cash for simplicity) that is posted frequently as a guarantee for due payments following mark to market, by the party to whom mark to market is negative. The guarantee is to be used by the party to whom mark to market is positive in case the other party defaults.

\item[Q:] That seems the end of counterparty risk then.

 \item[A:]Indeed, Collateral would be the main and most effective tool against counterparty risk, with two caveats. It is not always effective, even under frequent margining, and it can be expensive. It is shown in \cite{BrigoCapponiPalla} and \cite{brigoetalcollateral} that even very frequent margining may not be enough to fully protect from Counterparty Risk. The fact is that in extreme scenarios, the portfolio value may have moved a lot from the last margining date, even if this was a few moments ago. In \cite{BrigoCapponiPalla} an example is given, with Credit Default Swaps (CDS) as underlying instruments, where the default of the counterparty triggers an immediate jump in the underlying CDS by contagion, so that the collateral that was posted an instant earlier is not enough to cover the loss.

\item[Q:] Is this a rather abstract case?

\item[A:] I wouldn't think so, given what happened in 2008 after Lehman's default, and also keeping in mind that we had seven credit events on financials that happened in one month during the period going from September, 7 2008 to October, 8 2008, namely the credit events on Fannie Mae, Freddie Mac, Lehman Brothers, Washington Mutual, Landsbanki, Glitnir and Kaupthing.

\item[Q:] And what is re-hypothecation then?

\end{itemize}

\section{Re-hypothecation}

\begin{itemize}

 \item[A:]Re-hypothecation means that the collateral that has been received as a guarantee can be utilized as an investment or as further collateral. Suppose we are again in the example above with a bank and a corporate client. Suppose that at a margining date the mark to market of the portfolio is in favor of the bank, i.e. positive to the bank, so that the corporate client is posting collateral. If re-hypothecation is allowed, the bank is free to re-invest the collateral. Now suppose that there is an extreme movement in the market, such that the mark to market of the portfolio turns in favor of the corporate, and before the next collateral adjustment margining date arrives, the bank (and not the corporate) defaults.

\item[Q:] Uh-oh 

\item[A:] Indeed, "Uh-oh", but as I said don't take it personally.
The bank defaults while the mark to market of the portfolio is in favor of the corporate. Also, the bank had reinvested the collateral that had been posted by the corporate earlier. So the corporate client takes a double punishment: a loss of mark to market, and a loss of collateral.

\item[Q:] This sounds like a problem

 \item[A:]It is, and parts of the industry have made pressure to forbid re-hypothecation. While this is reasonable, the impossibility to re-invest collateral makes it particularly expensive, since the collateral taker needs to remunerate the interest on collateral to the collateral provider, now without the possibility to re-invest collateral.

\item[Q:] What is the extent of the impact of re-hypothecation on CVA?

 \item[A:]This has been studied in a few papers, see for example again \cite{BrigoCapponiPalla} and \cite{brigoetalcollateral}.

\item[Q:] So many things to read... what about Netting then?

\end{itemize}

\section{Netting}

\begin{itemize}

 \item[A:]Netting is an agreement where, upon default of your counterparty, you do not check the losses at single deal level but rather at the netted portfolio level.

\item[Q:] Could you provide an example please?

 \item[A:]Suppose you are the bank and you are trading two interest rate swaps with the same corporate, whose recovery rate is 0.4. Suppose at a point in time the two swaps have exactly the opposite value to the bank, say +1 M USD and -1M USD respectively.
    Now assume that the corporate client defaults. In the case with netting, the two swaps are netted, so that we compute 1M - 1M =0 and there is no loss to account for. In the case without netting, the two deals are treated separately. In the first swap, the bank loses (1-REC)1M = 0.6 M. In the second swap, the bank loses nothing.

\item[Q:] I see...

 \item[A:]Now in view of charging a fair CVA to the corporate, the bank needs to know whether there is netting or not, since as you have seen the difference can be rather important. In general unilateral CVA with netting is always smaller than without netting.

\item[Q:] And why is that?

\item[A:]  This is because CVA is like a call option with zero strike on the residual value of the deal, and an option on a sum is smaller than the sum of options.

\item[Q:] Has netting been studied?

\item[A:] There is a paper on netting for interest rate swaps where an approximate formula has also been derived, see \cite{BrigoMas}, but there is no wrong way risk. Netting with wrong way risk has been examined in \cite{BrigoPalla07}.

\end{itemize}

\section{Funding}

\begin{itemize}

\item[Q:] Ok, we covered quite a lot of stuff. There is a further topic I keep hearing around. It's the inclusion of Cost of Funding into the valuation framework. Is this actually happening?

 \item[A:]Yes, that's all the rage now. If you attend a practitioner conference, a lot of talks will be on consistent inclusion of funding costs. However, very few works try to build a consistent picture where funding costs are consistently included together with CVA, DVA, collateral, closeout etc.

\item[Q:] Some examples?

\item[A:] The working paper \cite{Crepey2011}, then published in \cite{Crepey2012a} and \cite{Crepey2012b}, is the most comprehensive treatment I have seen so far. The only limitation is that it does not allow for underlying credit instruments in the portfolio, and has possible issues with FX. It is a very technical paper. A related framework that is more general and includes most recent literature as a special case is in \cite{BrigoPallaPerini}. Earlier works are partial but still quite important.

\item[Q:] For example?

\item[A:] The industry paper \cite{Piter2010} considers the problem of replication of derivative transactions under collateralization but without default risk and in a purely classical Black and Scholes framework, considering then two relatively basic special cases. When you derive the first basic results that way, you have to be very careful on  the way you formulate the self financing condition. 



%



\item[Q:] I would imagine that default modeling is important in collateral and funding, isn't it? That is the reason why collateral is introduced in the first place.

\item[A:] Indeed, but inclusion of default risk is quite complicated. The fundamental funding implications in presence of default risk have been considered in simple settings first in \cite{MoriniPrampolini}, see also \cite{Castagna2011}. These works focus on particularly simple products, such as zero coupon bonds or loans, in order to highlight some essential features of funding costs.  \cite{Fujii2010} analyzes implications of currency risk for collateral modeling. \cite{Burgard2011} resorts to a PDE approach to funding costs. As I mentioned above \cite{Crepey2011}, then published in \cite{Crepey2012a} and \cite{Crepey2012b}, and \cite{BrigoPallaPerini} remain the most general treatments of funding costs to date. These papers show how complicated it is to formulate a proper general framework with collateral and funding but inclusive of default risk.

\item[Q:] What are the findings in Morini and Prampolini \cite{MoriniPrampolini}? I heard of this paper when it was still a preprint. 

\item[A:] One important point in Morini and Prampolini \cite{MoriniPrampolini} is that in simple payoffs such as bonds DVA can be interpreted as funding, in order to avoid double counting. However, this result does not extend to general payoffs, where different aspects interact in a more complex way and the general approach of Crepey \cite{Crepey2011}  or Pallavicini et al  \cite{BrigoPallaPerini} is needed.

\item[Q:] All right, ten more papers to read, but what is the funding problem, basically?

\item[A:] To put it in a nutshell, when you need to manage a trading position, you may need to obtain cash in order to do a number of operations: hedging the position, posting collateral, and so on. This is cash you may obtain from your Treasury department or in the market. You may also receive cash as a consequence of being in the position: a coupon, a notional reimbursement, a positive mark to market move, getting some collateral, a closeout payment. All such flows need to be remunerated: if you are borrowing, this will have a cost, and if you are lending, this will provide you with some revenues. Including the cost of funding into your valuation framework means to properly account for such features.

\item[Q:] Well looks like accounting to me

\item[A:] [Sighing] The trick is doing this consistently with all other aspects, especially counterparty risk. A number of practitioners advocate a ``Funding Valuation Adjustment", or FVA, that would be additive so that the total price of the portfolio would be

RISK FREE PRICE + DVA - CVA + FVA

However, it is not that simple. Proper inclusion of funding costs leads to a recursive pricing problem that may be formulated as a backwards stochastic differential equation (BSDE, as in \cite{Crepey2011}) or to  a discrete time backward induction equation (as in \cite{BrigoPallaPerini}). The simple additive structure above is not there in general.

\item[Q:] I doubt the banks will be willing to implement BSDEs, and I also doubt the regulators will prescribe that. We need something simple coming out of this.

\item[A:] All of a sudden you become reasonable and moderate? That's good [smiling]. However, sometimes it isn't possible to simplify dramatically.

\end{itemize}

\section{Hedging Counterparty Risk: CCDS}

\begin{itemize}

 \item[Q:]My last question is this. From what you said above, it looks like Basel III may impose quite some heavy capital requirements for CVA. Collateralization is a possible way out, but it may become expensive for some firms and lead to a liquidity strain, while firms that are not organized for posting collateral may be in troubles. \cite{watt} reports the case of the leading German airline: bear with me, I am low tech compared to you [Pulls out a pice of paper with part of an article]

\begin{quote}
"The airline's Cologne-based head of finance, Roland Kern, expects its earnings to become more volatile "not because of unpredictable passenger numbers, interest rates or jet fuel prices, but because it does not post collateral in its derivatives transactions".
\end{quote}

Indeed, without the possibility to post collateral, the firms would be subject to heavy CVA capital requirements. Is there a third way?

 \item[A:] There have been proposals for market instruments that can hedge CVA away, or reduce its capital requirements in principle. One such instrument, for example, is the Contingent Credit Default Swap (CCDS).
    
\item[Q:] What is a CCDS? Anything to do with standard CDS?

 \item[A:] It is similar to a CDS, but when the reference credit defaults, the protection seller pays protection on a notional that is not fixed but is rather given by the loss given default (1 - recovery) fraction of the residual value of a reference Portfolio at that time, if positive.
    
\item[Q:] So there is both a reference credit, against whose default protection is traded, and a reference portfolio?

 \item[A:] Consider this example. Suppose the Bank1 buys a contingent CDS, offering protection against default of her corporate client, which is the reference credit.  Protection is bought by the bank on the portfolio the bank is trading with the client. The bank buys this protection from another bank, say Bank2. The payoff of the default leg of the Contingent CDS to Bank1 is exactly the unilateral CVA Bank1 would measure against the corporate client on the traded portfolio. So if Bank2 is default-free, with the CCDS Bank1 is perfectly hedged against CVA on the reference portfolio traded with the corporate client, since the CVA payoff will be matched exactly by the CCDS protection leg.
    
\item[Q:] Have these products been popular in the past?

\item[A:] Not really. [Visualizes on the tablet the scan of a newspaper page]. The Financial Times was commenting back in 2008:

\begin{quotation}
 "[...]Rudimentary and idiosyncratic versions of these so-called CCDS have existed for five years, but they have been rarely traded due to high costs, low liquidity and limited scope. [...]
 Counterparty risk has become a particular concern in the markets for interest rate, currency, and commodity swaps - because these trades are not always backed by collateral.[...]
 Many of these institutions - such as hedge funds and companies that do not issue debt - are beyond the scope of cheaper and more liquid hedging tools such as normal CDS.
 The new CCDS was developed to target these institutions (Financial Times, April 10, 2008)."
\end{quotation}

Interest on CCDS has come back in 2011 now that CVA capital charges risk to become punitive. However, CCDS do not fully solve the problem of CVA capital requirements. First of all, there is no default free Bank2, so the CCDS itself would be subject to Counterparty risk. Also, it is not clear how CCDS would work in the bilateral case. And the hedging problem of a possible Bilateral CCDS (with all the DVA problems seen above) would fall on Bank2, so that the problem is only moved. While CCDS can be helpful in limited contexts, it is probably worth looking for alternatives. 

\item[Q:] So the market forgot about CCDS?

\item[A:] Not really. In fact, CCDS are now finally standardized on index portfolios by ISDA. ISDA came out with templates and documentation for CCDS, you may find those on the ISDA web site. Still, most of the problems I mentioned above are still there. This is prompting the industry to look for other solutions that may be effective also across several counterparties at the same time. 

\item[Q:] For example? 

\item[A:] CVA securitization could be considered, although the word ``securitization" is not much popular these days.

\item[Q:] Is there any proposed form of CVA restructuring, or securitization?

\end{itemize}

\section{Restructuring Credit Risk: CVA-CDOs and Margin Lending}

\begin{itemize}

\item[A:] [Concentrating, looking tired] There are a few. I am familiar with a few deals that have been discussed in the press, and in the Financial Times blog Alphaville in particular \cite{pollack2012}.

\item[Q:] The FT? Looks like this made the mainstream media

\item[A:] Yes. Let me show you [connects with the FT aplahville web site]. 

\medskip

"In short, Barclays has taken a pool of loans and securitised them, but retained all but the riskiest piece. On that riskiest Euro 300m, Barclays has bought protection from an outside investor, e.g. hedge fund. That investor will get paid coupons over time for their trouble, but will also be hit with any losses on the loans, up to the total amount of their investment. To ensure that the investor can actually absorb these losses, collateral is posted with Barclays."

\item[Q:] Looks like a CDO from the little I know? Looks like an equity tranche backed by collateral.  

\item[A:] Yes, collateral is key here. The blog continues: 

\medskip

"This point about collateral means that, at least in theory, Barclays is not exposed to the counterparty risk of the hedge fund. This is especially important because the hedge fund is outside the normal sphere of regulation, i.e. they aren't required to hold capital against risk-weighted assets in the way banks are."

\medskip

Notice this point of transferring risk outside the regulated system. This is a point that is stressed also in the OECD paper \cite{oecd}. The blog continues:

\medskip

"[...]  And then there is the over-engineering element whereby some deals were, and maybe still are, done where the premiums paid over time to the hedge fund are actually equal to or above the expected loss of the transaction. That the Fed and Basel Committee were concerned enough to issue guidance on this is noteworthy. It'll be down to individual national regulators to prevent "over-engineering", and some regulators are more hands-on than others."

So there you have it. 

\item[Q:] Interesting, are you aware of any other such deals? 

\item[A:] I know of a different one called SCORE. Again FT alphaville, this time from \cite{pollack2012b}:

\medskip

"RBS had a good go at securitising these exposures, but the deal didn't quite make it over the line. However, Euroweek reports that banks are still looking into it:

\medskip

{\indent{\emph{'Royal Bank of Scotland’s securitisation of counterparty credit risk, dubbed Score 2011, was pulled earlier this year, but other banks are said to be undeterred by the difficulties of the asset class, and are still looking at the market. However, other hedging options for counterparty risk may have dulled the economics of securitising this risk since the end of last year.'}}}

\medskip

So this is has not been that successful. 

\item[A:] Not really. The latest I heard of is Credit Suisse: \cite{pollack2012c}:

\medskip

"Last week Credit Suisse announced it had bought protection on the senior slice of its unusual employee compensation plan. The Swiss bank pays some of its senior bankers using a bond referencing counterparty risk, which also involves shifting some counterparty credit risk from the bank to its workers."

\medskip

So that is like buying protection from your own employees. Interesting concept if you think about it. That way the employee, in theory, is incentivized in improving the risk profile of the company.

\item[Q:] Maybe I'm a rookie, but to be honest I wouldn't be too happy if I were paid that kind of bonus. It may work for super-senior employees, like you, but for me... well... I don't participate into the important decisions of the company, I'm not a decision maker.

\item[A:] You overestimate my importance, I'm not the CEO, CFO, CRO, CIO, or C$*$O, I'm just your average risk manager!!

\item[Q:] But is this all about Counterparty Risk restructuring? No other idea? No new idea?

\item[A:] There are actually more innovative ideas. On CVA securitization, see for example \cite{albanese2011}, that advocates a global valuation model. The more model--agnostic \cite{albanesebrigooertel} explains how margin lending through quadri-partite or penta-partite structures involving clearing houses would be effective in establishing a third way.

\item[Q:] [Excitedly] Can you tell me more? This sounds intriguing.

\item[A:] Let me borrow from \cite{albanese2011} and \cite{albanesebrigooertel}, to which I refer for the full details. If I understood correctly, the structure is like this [draws Figure \ref{fig_abcd} on her tablet]

\begin{figure}[t]
\begin{center}
\includegraphics[width = 11cm]{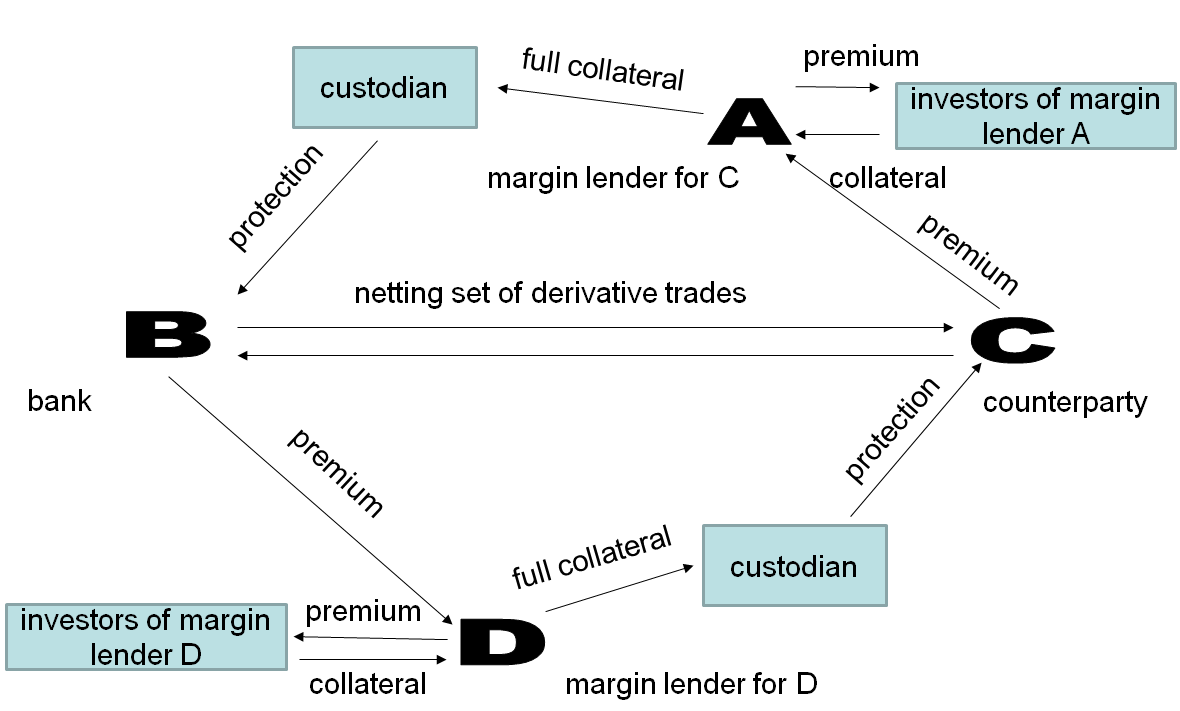}
\end{center}
\caption{General counterparty scheme including quadri-partite structure.}\label{fig_abcd}
\end{figure}

\item[Q:] How does this picture work?

\item[A:]
Traditionally, the CVA is typically charged by the structuring bank $B$ either on an upfront basis or it is built into the structure as a fixed coupon stream. The above deals we discussed, such as Papillon and Score, are probably of this type too. Margin lending instead is predicated on the notion of floating rate CVA payments with periodic resets...

\item[Q:] What is ``floating rate CVA"?

\item[A:]
Whichever formulation of CVA and DVA is chosen, once could postulate that CVA and DVA are paid periodically on rolling protection intervals. The related CVA is term "Floating Rate CVA" (FRCVA), and similarly for DVA.
Assume for simplicity that we are in a bi-partite transaction between the default-free bank B and the defaultable counterparty (say a corporate client) C. In principle, instead of charging CVA upfront at time 0 for the whole maturity of the portfolio, the bank may require a CVA payment at time 0 for protection on the exposure up to 6 months. Then in 6 months the bank will require a CVA payment for protection for further six months on what will be the exposure up to one year, and on and on, up to the final maturity of the deal. Such a CVA would be an example of FRCVA.

\item[Q:] Ok, back to Figure \ref{fig_abcd}

\item[A:]
I was saying that margin lending is based on the notion of floating rate CVA payments with periodic resets,
 and is designed in such a way to transfer the conditional credit spread volatility risk and the mark-to-market volatility risk from the bank to the counterparties.
We may explain this more in detail by following the arrows in the Figure.

\item[Q:] Ok, I'm ready, looking at Figure \ref{fig_abcd} [Excited]

\item[A:] Relax a second. The counterparty C, a corporate client, has problems with posting collateral periodically in order to trade derivatives with bank B. To avoid posting collateral, C enters into a margin lending transaction. C pays periodically (say semi-annually) a floating rate CVA to the margin lender A (`premium' arrow connecting C to A), which the margin lender A pays to investors (premium arrow connecting A to Investors). This latest payment can have a seniority structure similar to that of a cash CDO.

\item[Q:] Dangerous territory there... [Grinning]

\item[A:] [Flashing an irritated look] Let me finish. In exchange for this premium, for six months the investors provide the margin lender A with daily collateral posting (`collateral' arrow connecting Investors to A) and A passes the collateral to a custodian (`collateral' arrow connecting A to the custodian). This way, if C defaults within the semi-annual period, the collateral is paid to B to provide protection (`protection' arrow connecting the custodian to B) and the loss in taken by the Investors who provided the collateral.

\item[Q:] Ok, so far it's clear.

\item[A:] At the end of the six months period, the margin lender may decide whether to continue with the deal or to back off. With this mechanism C is bearing the CVA volatility risk, whereas B is not exposed to CVA volatility risk, which is the opposite of what happens with traditional upfront CVA charges.

\item[Q:] So one of the big differences with traditional CVA is that in this structure the CVA volatility stays with the counterparty C that is generating it, and does not go to the bank B.

%

 \item[A:] Indeed, \cite{albanesebrigooertel} argue that whenever an entity's credit worsens, it receives a subsidy from its counterparties in the form of a DVA positive mark to market which can be monetized by the entity's bond holders only upon their own default. Whenever an entity's credit improves instead, it is effectively taxed as its DVA depreciates. Wealth is thus transferred from the equity holders of successful companies to the bond holders of failing ones, the transfer being mediated by banks acting as financial intermediaries and implementing the traditional CVA/DVA mechanics.

\item[Q:] Whoa!

\item[A:] [Smiling] It's good to see someone still so refreshingly enthusiastic. Rewarding failing firms with a cash subsidy may be a practice of debatable merit as it skews competition. But rewarding failing firms with a DVA benefit is without question suboptimal from an economic standpoint: the DVA benefit they receive is paid in cash from their counterparties but, once received in this form, it cannot be invested and can only be monetized by bond holders upon default.

\item[Q:] I see...

\item[A:] Again, \cite{albanesebrigooertel} submits that margin lending structures may help reversing the macroeconomic effect by eliminating long term counterparty credit risk insurance and avoiding the wealth transfer that benefits the bond holders of defaulted entities.

\item[Q:] I can see a number of problems with this. First, proper valuation and hedging of this to the investor who are providing collateral to the lender is going to be tough. I recall there is no satisfactory standard for even simple synthetic CDOs. One would need an improved methodology. 

\item[A:] Weren't you the one complaining about the situation being already too complicated? But indeed, the modeling problems have been highlighted for example in \cite{brigopallatorre}.  Admittedly this requires and effective global valuation framework, see for example the discussion in \cite{albanese2011}.

\item[Q:] The other problem is: what if all margin lenders pull off at some point due to a systemic crisis?

\item[A:] That would be a problem, indeed, but \cite{albanesebrigooertel} submits that the market is less likely to arrive in such a situation in the first place if the wrong incentives to defaulting firms are stopped and an opposite structure, such as the one in Figure \ref{fig_abcd} is implemented. There is also a penta-partite version including a clearing house.

\item[Q:] Mmmm.... I understand that if the counterparty credit risk deteriorates, the counterparty will be charged more. Isn't this something that could compromise the relationship of a bank with an important client?

\item[A:] It is. One could diminish this by putting a cap and a floor on the floating CVA, and then however pricing and hedging this cap risk would go back to part of the original problems. Part of the volatility would be still neutralized, however.

\item[Q:] But I can see the appeal of floating CVA. It's like car insurance. If you drive well next year you expect to be rewarded with less premium next year. If you drive poorly and have an accident you may expect your premium to go up. Everyone accepts this. So I think it could work also with banks.

\item[A:] Well the client relationship is more complex with banks, but yes that is an initial analogy. In any case there is much more work to do to assess this framework properly, and it is evolving continually. 

\item[Q:] This looks like a good place to stop then.

\item[A:] Indeed. [Smiling but looking tired]

\item[Q:] Thanks for your time and patience. [Smiling gratefully but still a little puzzled]

\end{itemize}

\end{document}